\documentclass[a4paper]{spie}  
 
\usepackage{amsmath,amsfonts,amssymb}
\usepackage{graphicx}
\usepackage[colorlinks=true, allcolors=blue]{hyperref}
\usepackage{multirow}

\title{A numerical simulation study of an astrometry case for MORFEO at the ELT}

\author[a,e]{Carmelo Arcidiacono}
\author[b,e]{Elisa Portaluri}
\author[a]{Marco Gullieuszik}
\author[b]{Michele Cantiello}
\author[c]{Francesca Annibali}
\author[c]{Paolo Ciliegi}
\author[a]{Matteo Simioni}
\author[a,e]{Daniela Fantinel}
\author[d,e]{Guido Agapito}
\author[a,e]{Demetrio Magrin}

\affil[a]{INAF - Osservatorio Astronomico di Padova (Italy)}
\affil[b]{INAF - Osservatorio Astronomico d'Abruzzo (Italy)}
\affil[c]{INAF - Osservatorio di Astrofisica e Scienza dello Spazio (Italy)}
\affil[d]{INAF - Osservatorio Astrofisico di Arcetri (Italy)}
\affil[e]{ADONI – Laboratorio Nazionale di Ottica Adattiva (Italy)}

\authorinfo{Further author information: (Send correspondence to C.A.)\\C.A.: E-mail: carmelo.arcidiacono@inaf.it, Telephone: +39 049 829 3414\\  E.P.: E-mail: elisa.portaluri@inaf.it, Telephone: +39 0862 412900}

\begin{document} 
\maketitle

\begin{abstract}
We report results from numerical simulations assessing astrometry measurements with the Multiconjugate Adaptive Optics Relay for ELT Observations (MORFEO) instrument on the Extremely Large Telescope (ELT). Using the Advanced Exposure Time Calculator (AETC), we evaluate MORFEO astrometric accuracy in moderately crowded fields. Our simulations account for spatially variable Point Spread Function (PSF), geometric distortion, and rotation-dependent variations. We computed focal plane coordinates using observed stellar distribution and computed population synthesis with the SPISEA tool, generating stellar magnitude distributions for MICADO filters at selected metallicities and stellar ages. Our analysis shows that MORFEO can achieve high-precision astrometry in the galaxy neighborhood (within $\mu < 24$ mag) by minimizing PSF enlargement and optimizing calibration strategies. These results inform future observational campaigns and contribute to the development of astrometric science cases for the ELT.

\end{abstract}


\section{INTRODUCTION}
\label{sec:intro}  

We report results from numerical simulations that assess astrometry measurements with the Multiconjugate Adaptive Optics Relay for ELT Observations (MORFEO\cite{ciliegiMAORYAdaptiveOptics2020a,ciliegiSystemOverview2021,ciliegiMAORYMORFEOELT2022a} instrument on the Extremely Large Telescope (ELT). Astrometry is the primary science case for MICADO\cite{daviesMICADOEELTAdaptive2010,daviesMICADOFirstLight2016}, the MORFEO client instrument. Using Advanced Exposure Time Calculator (AETC)\cite{falomoAETCAdvancedExposure2011}, we evaluated the astrometric accuracy of MORFEO in moderately crowded fields. Our findings reveal MORFEO potential for ELT astrometry and provide insights for future observational campaigns. 

We computed focal plane coordinates by applying observed stellar distribution and computed population synthesis, We used the SPISEA\cite{hosekSPISEAPythonbasedSimple2020} python tool to compute stellar magnitude distributions in specific filters (MICADO filters) for selected metallicity and stellar ages. Using as a reference the colour magnitude diagram generated by SPISEA of a selected astronomical target we applied a spatial distribution mimicking the observed one and a set of population ages.

MORFEO employs state-of-the-art adaptive optics technologies, including multiple laser guide stars, multiple deformable mirrors, and low-noise detectors following the Multi-Conjugate Adaptive Optics approach\cite{beckersIncreasingSizeIsoplanatic1988,beckersDetailedCompensationAtmospheric1989}. These advancements will enhance near-infrared observations with MICADO, offering high spatial resolution (usually below 20 mas), uniformly corrected large field of view ($\approx 50"$), and high sky coverage ($> 50$)\% for South Galactic Pole star density) and exceptional astrometric accuracy ($< 50–150$ microarcseconds), all paired with extreme sensitivity comparable to the James Webb Space Telescope (JWST).

Using an optimal set of reference sources in the field of the science targets, the low-order astrometric calibrations carried out by MICADO allow us to meet the astrometric accuracy requirement. The optimal setup for high astrometry performance involves limiting the integration time of each exposure to 30 seconds for broadband filters and using the H band, minimising the effect of different colour refraction between WFS and Science Camera. MICADO's astrometric mode utilizes 'zoom' mode optics, providing a pixel scale of 1.5 mas/px.

Our simulations do not account for the residual of the MCAO correction. The photon and detector noise affects the low-order corrections provided sensing the Natural Guide Stars (NGS), which may significantly impact astrometric stability and differential astrometry performance. Additionally, the influence of field rotation with a fixed pattern of LGS stars could cause time variations in the radial correction terms. 

The current operation scheme for astrometry observations foresees the determination of the low-order optical geometric distortions of the camera for each frame using high signal-to-noise ratio stars, with high-order distortions evaluated through a calibration pinhole mask placed on the intermediate focal plane just ahead of MICADO in the MORFEO path.

As a summary of the requirement, MORFEO will not enlarge the PSF more than 1/10 of Full Width at Halh Maximum (FWHM) in the worst case of elevation 85 degrees, for a max exposure time of 30 seconds and 2 minutes in the case of broad and narrow band, respectively.

In this paper, we take as reference the blue compact dwarf NGC 1705, D=5.1 Mpc with the aim of establishing the range of distance and star brightness are open to MORFEO-MICADO investigation. We took the following approach: simulating the (relative) positions of the stars inside the clusters at varying SNR, fitting coefficients of the distortion polynomials on a bright sub-sample of stars and apply this (in a removal sense) to the whole set of measured coordinates.

Our analysis demonstrates the feasibility of high-precision astrometry in the galaxy neighbourhood within $\mu < 24$ mag, highlighting the importance of the calibration strategy and optimal PSF fitting.

\section{MORFEO-MICADO SETUP FOR ASTROMETRY}
\label{sec:setup}

\begin{figure} [ht]
   \begin{center}
   \begin{tabular}{c} 
   \includegraphics[height=7cm]{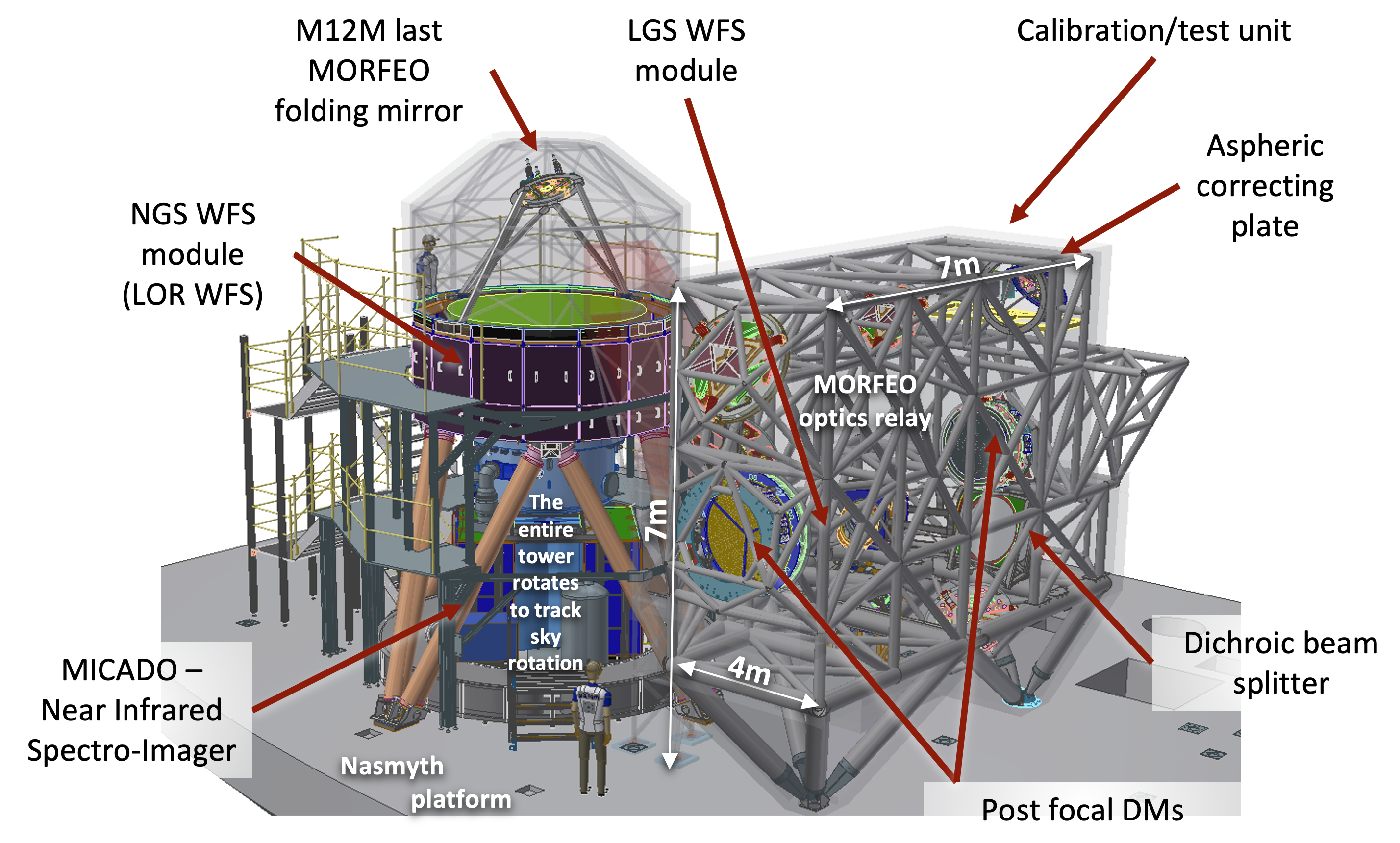}
   \end{tabular}
   \end{center}
   \caption[MORFEO-MICADO Setup]  
   {Overview of the MORFEO and MICADO assembly on the ELT Nasmyth-B platform. \label{fig:setup} }
   \end{figure}

MORFEO senses optical turbulence in the atmospheric volume above the telescope using six sodium layer beacons provided by the ELT and three Natural Guide Stars (NGS) within an 80" radius FoV. The Low Order (LO) and Reference (R) Wavefront Sensors (WFS) of MORFEO sense in H (LO) and R bands, respectively. The optimal AO performance can be obtained using a constellation as shown in Figure~\ref{fig:setup}, using stars brighter than H=16-17. MORFEO feeds the MICADO spectro-imaging camera working from I to Ks band.

\begin{figure} [ht]
   \begin{center}
   \begin{tabular}{c} 
   \includegraphics[height=8cm]{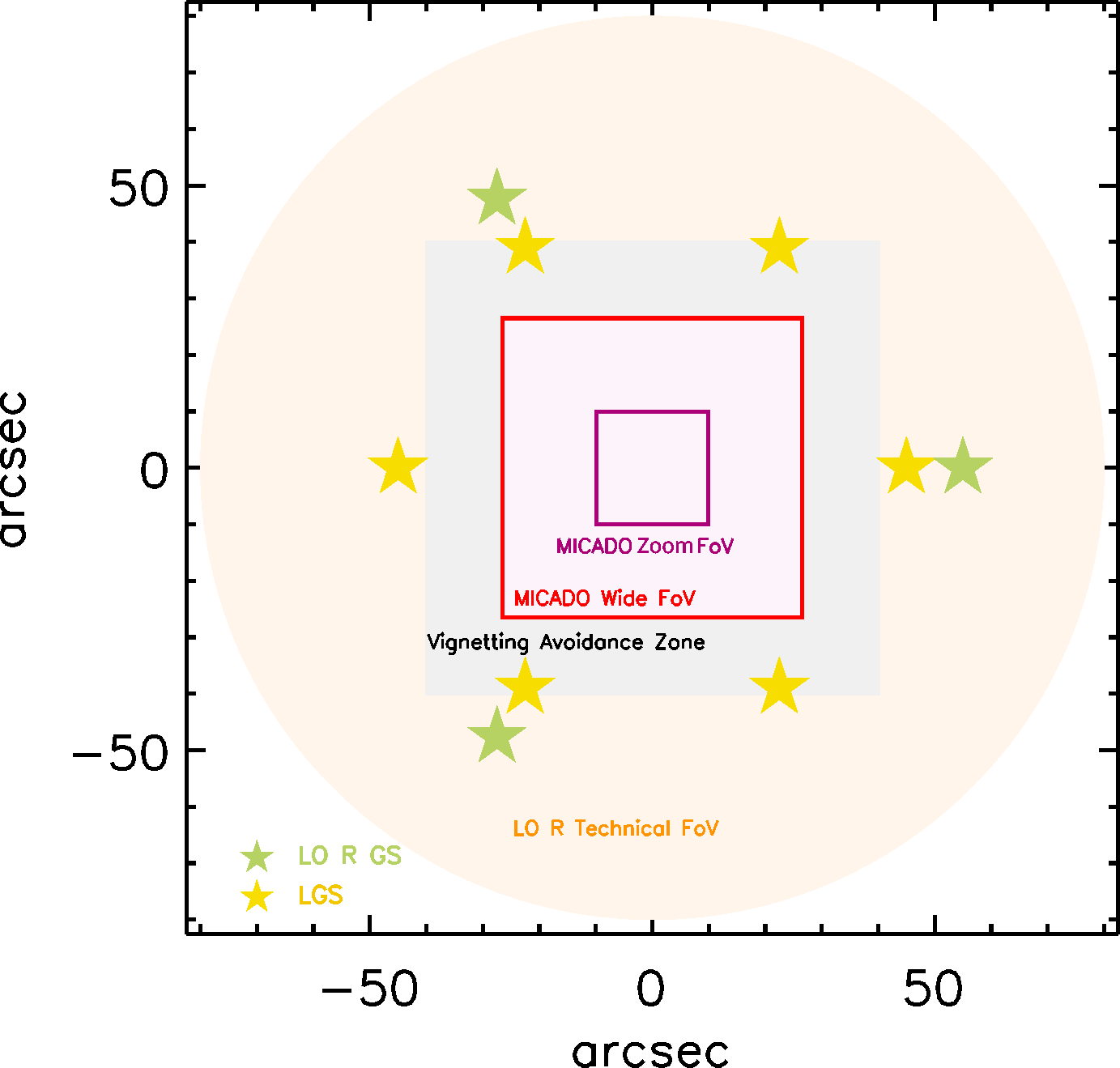}
   \end{tabular}
   \end{center}
   \caption[MICADO Field of View]  
   {Technical and MICADO FoV, the preferred astrometry mode FoV is the magenta box with size 20” x 20” (1.5mas/px). \label{fig:FoV} }
   \end{figure}

The preferred astrometry mode FoV for MICADO is 20" x 20" with a pixel scale of 1.5 mas/px (zoom mode optics), as shown in Figure~\ref{fig:FoV}.

\section{A SCIENTIFIC EXAMPLE FOR ASTROMETRY}
\label{sec:science}

\begin{figure} [ht]
   \begin{center}
   \begin{tabular}{c} 
   \includegraphics[height=9cm]{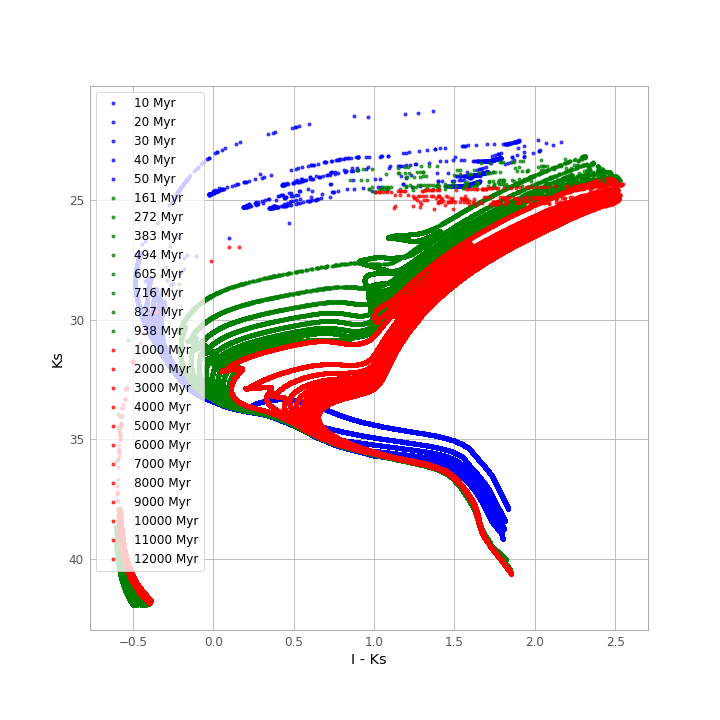}
   \end{tabular}
   \end{center}
   \caption[Color Magnitude Diagram]  
   {Colour Magnitude Diagram for the different stellar populations composing NGC 1705. \label{fig:cmd} }
   \end{figure}

\begin{figure} [ht]
   \begin{center}
   \begin{tabular}{c} 
   \includegraphics[height=7cm]{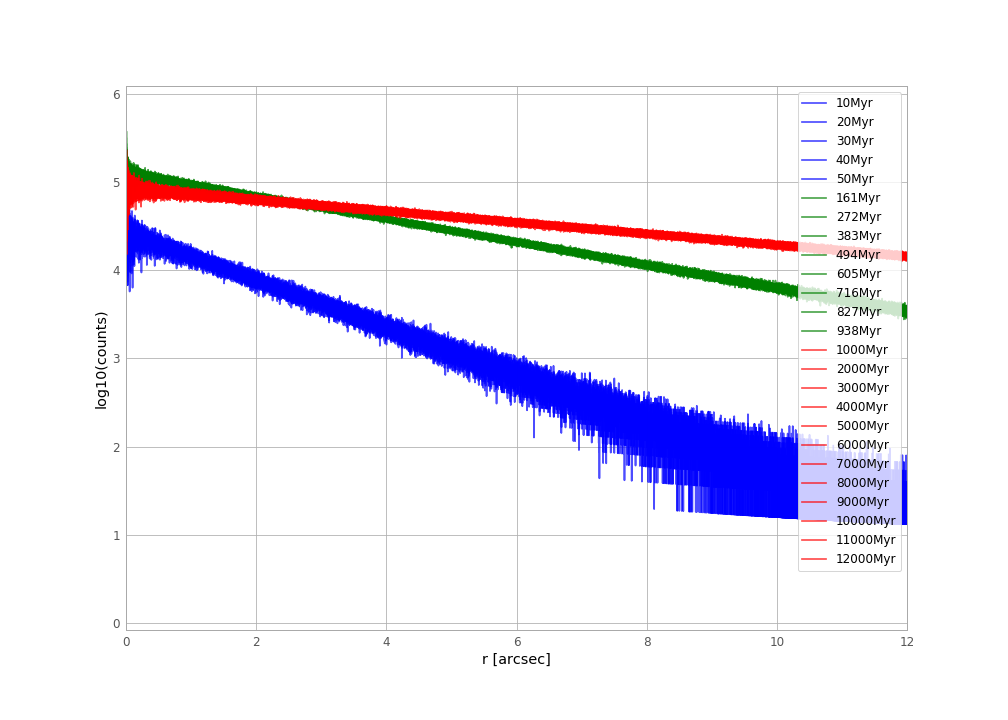}
   \end{tabular}
   \end{center}
   \caption[Spatial Distribution]  
   {Spatial density distributions mimicking NGC 1705. \label{fig:spatial_distribution} }
   \end{figure}

To start the analysis, we simulated a stellar field with appropriate magnitude values and realistic coordinate positions. We used SPISEA to generate the colour magnitude diagram (CMD) for the blue compact dwarf NGC 1705 at a distance of 5.1 Mpc. The CMD accounts for metallicity and stellar ages (Figure~\ref{fig:cmd}). We applied the spatial distribution that mimics the observed radial profiles (Figure~\ref{fig:spatial_distribution}).

\subsection{Verifying available software}
\label{sec:sw}

\begin{figure} [ht]
   \begin{center}
   \begin{tabular}{c} 
   \includegraphics[height=7cm]{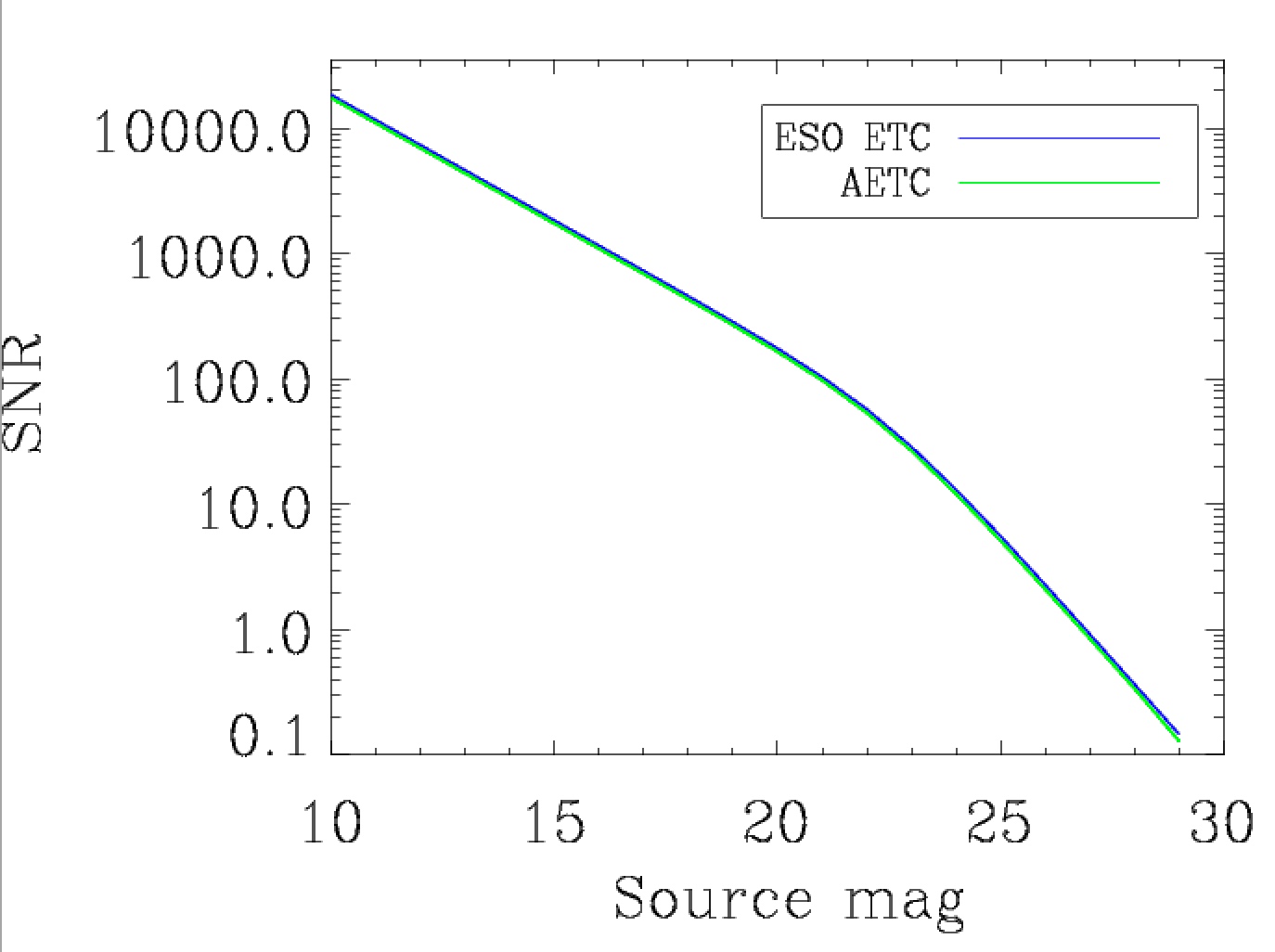}
   \end{tabular}
   \end{center}
   \caption[SNR Comparison]  
   {Comparison of the SNR curves derived using ESO ETC and AETC for a generic camera with 5mas/px sampling. \label{fig:SNR} }
   \end{figure}

We simulated the MORFEO-MICADO performance in terms of Signal to Noise Ratio (SNR) using the AETC \href{http://aetc.oapd.inaf.it}{http://aetc.oapd.inaf.it}. The tool provides SNR vs magnitude computation starting from ELT, MORFEO, MICADO, and sky information. It was verified using the ESO ELT Exposure Time Calculator (ETC). Figure~\ref{fig:SNR} compares the SNR curves derived using ESO ETC \href{https://www.eso.org/observing/etc}{https://www.eso.org/observing/etc} and AETC  for a generic camera with 5 mas/px sampling, in the most similar conditions adapting the AETC configuration to the Laser Tomography case in the ESO-ETC.

Nevertheless, the PSF used by the two tools are different, the comparison of the SNR shows a nice agreement of the Exposure Time Calculator for the ESO-ELT and the AETC, see Fig.\ref{fig:result}. In the case of ESO-ELT with H band and 5 mas / px, encircled energy (EE) = 6.5\%, while for AETC we measured directly from the simulated images EE=7.5\%.  The AETC was configured to work with good-performance PSF corresponding to the optimal configuration of the NGS asterism (as the example in Figure\ref{fig:FoV}) and for the median seeing conditions (seeing 0.67 arcsec).
\section{Performance Estimation}
\begin{figure}
    \centering
    \includegraphics[height=10cm]{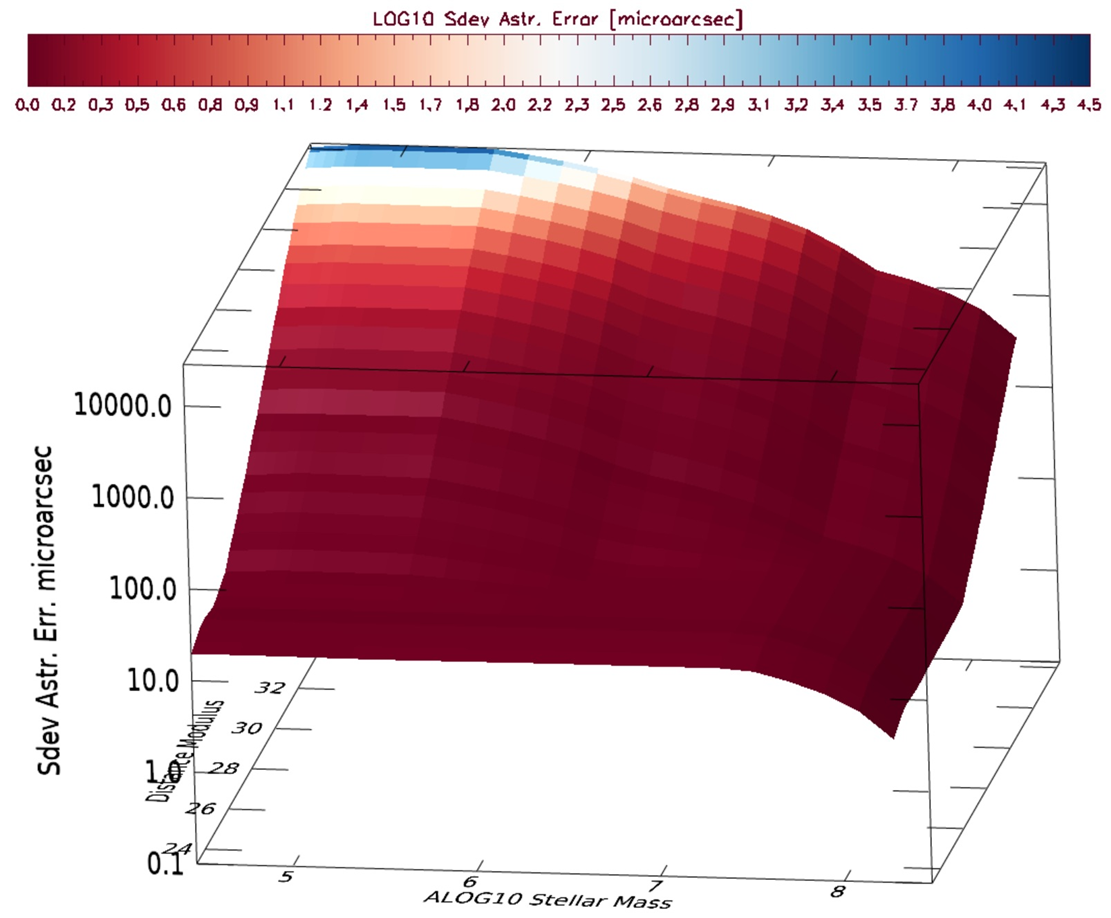}
    \caption{The surface plot gives the error, standard deviation, in the positioning of the high SNR stars available in the 20”x20” MICADO FoV.}
    \label{fig:result}
\end{figure}

We focus on the achievable centroiding error and translate this into an error in the determination of low-order distortions to be estimated from the image. 
We perform the propagation into a positional centroid error by coupling to the SNR of the simulated stars onto an error on the centroid through 
\begin{equation}
\sigma_{centroid} = 1/\pi (\lambda/D)/\sqrt{N} \approx 1/\pi {FWHM}/{SNR}
\end{equation}
as in reference\cite{lindegrenPhotoelectricAstrometryComparison1978}, where $N$ is the number of photo-electrons detected from the source in the aperture area, here the expected Full Width at Half Maximum $FWHM$. As explained above, once in operation, in the MICADO astrometry mode, we plan to correct for very low-order geometric distortions as derived from the positions of a set of bright reference stars. In this analysis we considered the centroid positions of the 100 brightest stars and compute the coefficients of the third order polynomial that makes the match between a set of measured star position and the original derived from NGC 1705, see also references\cite{arcidiaconoMAORYSciencePreparation2020,arcidiaconoMORFEOOperationalConcept2022}. The expectations is to compute coefficients very close to zero for all the cases with high SNR.
Now we can apply the estimated astrometric distortion polynomials to the whole set of simulated stars. As merit function, we computed the standard deviation of the position error for the stars with $SNR > 300$ or, if the SNR is too low, onto the first 300 brighter stars. We used as references for the calculation the spatial x, y coordinates and H magnitude distributions generated for NGC 1705 (distance modulus, $\mu =28.5$ mag, and total mass $3\times10^8$M\textsubscript{\(\odot\)}) to create further cases at different stellar masses ($10^4$-$10^8$) and distance moduli (24-34). In Figure\ref{fig:result}, we show the results.

We show that limiting the error sources for centroiding accuracy only to the PSF centroid-SNR, the astrometric error is below 10$\mu as$ for distance moduli $\mu<24$ mag, in the whole explored range of masses from 10$^4$ to 10$^8$ M\textsubscript{\(\odot\)}.

\section{Conclusions}
\label{sec:conc}

Our numerical simulations using AETC demonstrate the potential of the MORFEO-MICADO system for high-precision astrometry on the ELT. The results are promising, showing that with proper calibration and optimization, MORFEO can achieve the astrometric accuracy needed for scientific studies of stellar populations, exoplanet detection, and dark matter mapping.

Future work will involve increasing the complexity of the simulated fields to include effects of blending and higher background typical of crowded environments, such as the cores of globular clusters. We also plan to integrate the simulations with actual observational data from existing telescopes to further validate and refine the models. These efforts will contribute to the development of MORFEO and guide future observational campaigns with the ELT.

 \acknowledgments 
We thank the teams involved in the development of AETC for their support and for providing the necessary tools for our simulations. The authors also acknowledge the contributions of the MICADO and MORFEO consortia and the ESO-ELT project team.

\bibliography{better} 

\begin{thebibliography}{10}

\bibitem{ciliegiMAORYAdaptiveOptics2020a}
Ciliegi, P., Agapito, G., Aliverti, M., Arcidiacono, C., Balestra, A., Baruffolo, A., Bergomi, M., Bianco, A., Bonaglia, M., Busoni, L., Cantiello, M., Cascone, E., Chinellato, S., Cianniello, V., Correia, J.-J., Cosentino, G., Caprio, V.~D., Devaney, N., Antonio, I.~D., Cianno, A.~D., Giammatteo, U.~D., Rico, G.~D., Dolci, M., Eredia, C., Farinato, J., Esposito, S., Fantinel, D., Feautrier, P., Foppiani, I., Giro, E., Gluck, L., Goncharov, A., Grani, P., Gullieuszik, M., Haguenauer, P., Henault, F., Louarn, M.~L., Magrin, D., Malone, D., Marafatto, L., Munari, M., Oberti, S., Pariani, G., Pettazzi, L., Plantet, C., Portaluri, E., Puglisi, A., Rabou, P., Ragazzoni, R., Rakich, A., Redaelli, E., Riva, M., Rochat, S., Rodeghiero, G., Salasnich, B., Sordo, R., Sztefek, M.-H., Valentini, A., Verinaud, C., Xompero, M., and Zoltan, H., ``{{MAORY}}: The adaptive optics module for the {{Extremely Large Telescope}} ({{ELT}}),'' in [{\em Adaptive {{Optics Systems VII}}}{\nolinebreak\hspace{0.1em}]},   {\bf 11448},
  205--212, SPIE (Dec. 2020).

\bibitem{ciliegiSystemOverview2021}
Ciliegi, P., Agapito, G., Aliverti, M., Arcidiacono, C., Baruffolo, A., Bonaglia, M., Busoni, L., Cascone, E., Caprio, V.~D., Devaney, N., Giammatteo, U.~D., Rico, G.~D., Esposito, S., Farinato, J., Feautrier, P., Foppiani, I., Giro, E., Goncharov, A., Hubert, Z., Magrin, D., Oberti, S., Ragazzoni, R., Redaelli, E., Riva, M., Rodeghiero, G., Salansich, B., and Xompero, M., ``System {{Overview}},'' Preliminary {{Design Review}} E-MAO-000-INA-DER-001, INAF (Jan. 2021).

\bibitem{ciliegiMAORYMORFEOELT2022a}
Ciliegi, P., Agapito, G., Aliverti, M., Annibali, F., Arcidiacono, C., Azzaroli, N., Balestra, A., Baronchelli, I., Baruffolo, A., Bergomi, M., Bianco, A., Bonaglia, M., Briguglio, R., Busoni, L., Cantiello, M., Capasso, G., Carl{\`a}, G., Carolo, E., Cascone, E., Chinellato, S., Cianniello, V., Colapietro, M., Correia, J.-J., Cosentino, G., D'Auria, D., Caprio, V.~D., Devaney, N., Antonio, I.~D., Cianno, A.~D., Dato, A.~D., Giammatteo, U.~D., Rico, G.~D., Dolci, M., Eredia, C., Esposito, S., Fantinel, D., Farinato, J., Feautrier, P., Foppiani, I., Genoni, M., Giro, E., Gluck, L., Goncharov, A., Grani, P., Greggio, D., Guieu, S., Gullieuszik, M., Haguenauer, P., Hubert, Z., Lapucci, T., Laudisio, F., Louarn, M.~L., Magrin, D., Malone, D., Marafatto, L., Munari, M., Oberti, S., Pariani, G., Pettazzi, L., Plantet, C., Portaluri, E., Puglisi, A., Rabou, P., Ragazzoni, R., Redaelli, E., Riva, M., Rochat, S., Rodeghiero, G., Salasnich, B., Savarese, S., Scalera, M., Schipani, P., Sordo, R., Sztefek, M.-H.,
  Valentini, A., and Xompero, M., ``{{MAORY}}/{{MORFEO}} at {{ELT}}: General overview up to the preliminary design and a look towards the final design,'' in [{\em Adaptive {{Optics Systems VIII}}}{\nolinebreak\hspace{0.1em}]},   {\bf 12185},  325--334, SPIE (Aug. 2022).

\bibitem{daviesMICADOEELTAdaptive2010}
Davies, R., Ageorges, N., Barl, L., Bedin, L.~R., Bender, R., Bernardi, P., Chapron, F., Clenet, Y., Deep, A., Deul, E., Drost, M., Eisenhauer, F., Falomo, R., Fiorentino, G., Schreiber, N. M.~F., Gendron, E., Genzel, R., Gratadour, D., Greggio, L., Grupp, F., Held, E., Herbst, T., Hess, H.-J., Hubert, Z., Jahnke, K., Kuijken, K., Lutz, D., Magrin, D., Muschielok, B., Navarro, R., Noyola, E., Paumard, T., Piotto, G., Ragazzoni, R., Renzini, A., Rousset, G., Rix, H.-W., Saglia, R., Tacconi, L., Thiel, M., Tolstoy, E., Trippe, S., Tromp, N., Valentijn, E.~A., Kleijn, G.~V., and Wegner, M., ``{{MICADO}}: The {{E-ELT}} adaptive optics imaging camera,'' in [{\em Ground-Based and {{Airborne Instrumentation}} for {{Astronomy III}}}{\nolinebreak\hspace{0.1em}]},   {\bf 7735},  900--911, SPIE (July 2010).

\bibitem{daviesMICADOFirstLight2016}
Davies, R., Schubert, J., Hartl, M., Alves, J., Cl{\'e}net, Y., {Lang-Bardl}, F., Nicklas, H., Pott, J.-U., Ragazzoni, R., Tolstoy, E., Agocs, T., {Anwand-Heerwart}, H., Barboza, S., Baudoz, P., Bender, R., Bizenberger, P., Boccaletti, A., Boland, W., Bonifacio, P., Briegel, F., Buey, T., Chapron, F., Cohen, M., Czoske, O., Dreizler, S., Falomo, R., Feautrier, P., Schreiber, N.~F., Gendron, E., Genzel, R., Gl{\"u}ck, M., Gratadour, D., Greimel, R., Grupp, F., H{\"a}user, M., Haug, M., Hennawi, J., Hess, H.~J., H{\"o}rmann, V., Hofferbert, R., Hopp, U., Hubert, Z., Ives, D., Kausch, W., Kerber, F., Kravcar, H., Kuijken, K., {Lang-Bardl}, F., Leitzinger, M., Leschinski, K., Massari, D., Mei, S., Merlin, F., Mohr, L., Monna, A., M{\"u}ller, F., Navarro, R., Plattner, M., Przybilla, N., Ramlau, R., Ramsay, S., Ratzka, T., Rhode, P., Richter, J., Rix, H.-W., Rodeghiero, G., Rohloff, R.-R., Rousset, G., Ruddenklau, R., Schaffenroth, V., Schlichter, J., Sevin, A., Stuik, R., Sturm, E., Thomas, J., Tromp, N.,
  Turatto, M., {Verdoes-Kleijn}, G., Vidal, F., Wagner, R., Wegner, M., Zeilinger, W., Ziegler, B., and Zins, G., ``{{MICADO}}: First light imager for the {{E-ELT}},'' in [{\em Ground-Based and {{Airborne Instrumentation}} for {{Astronomy VI}}}{\nolinebreak\hspace{0.1em}]},   {\bf 9908},  99081Z, {International Society for Optics and Photonics} (Aug. 2016).

\bibitem{falomoAETCAdvancedExposure2011}
Falomo, R., Fantinel, D., and Uslenghi, M., ``{{AETC}}: {{Advanced Exposure Time Calculator}},'' in [{\em Applications of {{Digital Image Processing XXXIV}}}{\nolinebreak\hspace{0.1em}]},   {\bf 8135},  813523, {International Society for Optics and Photonics} (Sept. 2011).

\bibitem{hosekSPISEAPythonbasedSimple2020}
Hosek, Jr., M.~W., Lu, J.~R., Lam, C.~Y., Gautam, A.~K., Lockhart, K.~E., Kim, D., and Jia, S., ``{{SPISEA}}: {{A Python-based Simple Stellar Population Synthesis Code}} for {{Star Clusters}},'' {\em The Astronomical Journal}~{\bf 160},  143 (Sept. 2020).

\bibitem{beckersIncreasingSizeIsoplanatic1988}
Beckers, J.~M., ``Increasing the size of the isoplanatic patch with multiconjugate adaptive optics.,'' in [{\em {{ESO Conference}} on {{Very Large Telescopes}} and Their {{Instrumentation}}}{\nolinebreak\hspace{0.1em}]},   {\bf 2},  693--703 (1988).

\bibitem{beckersDetailedCompensationAtmospheric1989}
Beckers, J.~M., ``Detailed compensation of atmospheric seeing using multiconjugate adaptive optics.,'' in [{\em Active {{Telescope Systems}}}{\nolinebreak\hspace{0.1em}]},  {\em Proc. {{SPIE}}} {\bf 1114},  215--217 (1989).

\bibitem{lindegrenPhotoelectricAstrometryComparison1978}
Lindegren, L., ``Photoelectric astrometry - {{A}} comparison of methods for precise image location,'' {\em International Astronomical Union Colloquium}~{\bf -1},  197--217 (Sept. 1978).

\bibitem{arcidiaconoMAORYSciencePreparation2020}
Arcidiacono, C., Portaluri, E., Cantiello, M., Gullieuszik, M., Ciliegi, P., Simioni, M., Marasco, A., and Dall'Ora, M., ``{{MAORY}} science preparation status,'' in [{\em Adaptive {{Optics Systems VII}}}{\nolinebreak\hspace{0.1em}]},   {\bf 11448},  625--635, SPIE (Dec. 2020).

\bibitem{arcidiaconoMORFEOOperationalConcept2022}
Arcidiacono, C., Busoni, L., Foppiani, I., Oberti, S., Agapito, G., Plantet, C., Salasnich, B., Balestra, A., Baruffolo, A., Portaluri, E., and Riva, M., ``{{MORFEO Operational Concept Description}},'' Preliminary {{Design Review}} E-MAO-000-INA-MAN-002, INAF (Sept. 2022).

\end{thebibliography}
\bibliographystyle{spiebib} 

\end{document}